\documentclass[12pt,preprint]{aastex}

\begin{document}

\def\arcsecpoint{$''\!.$}
\def\arcminpoint{$'\!.$}
\def\deg{$^{\rm o}$}
\def\ltsim{\raisebox{-.5ex}{$\;\stackrel{<}{\sim}\;$}}
\def\gtsim{\raisebox{-.5ex}{$\;\stackrel{>}{\sim}\;$}}



\shortauthors{Dunn, et al.}
\shorttitle{Internet Database of UV Seyfert Lightcurves}

\title{An Internet Database of Ultraviolet Lightcurves for Seyfert Galaxies}

\author{Jay P. Dunn\altaffilmark{1,2},
Brian Jackson\altaffilmark{3},
Rajesh P. Deo\altaffilmark{1}
Chris Farrington\altaffilmark{1},
Varendra Das\altaffilmark{1},
\& D. Michael Crenshaw\altaffilmark{1}}

\altaffiltext{1}{Department of Physics and Astronomy, Georgia State University,
Atlanta, GA 30303}

\altaffiltext{2}{Email: dunn@chara.gsu.edu}

\altaffiltext{3}{Lunar \& Planetary Lab, University of Arizona, Tuscon, AZ
85721}

\begin{abstract}

Using the Multimission Archives at Space Telescope (MAST), we have extracted 
spectra and determined continuum light curves for 175 Seyfert Galaxies that
have been observed with the {\it International Ultraviolet Explorer (IUE)}
and the Faint Object Spectrograph (FOS) on the {\it Hubble Space Telescope
(HST)}. To obtain the light curves as a function of Julian Date, we used fix bins
in the object's rest frame, and measured small regions (between 30 and 60 \AA)
of each spectrum's continuum flux in the range 1150 \AA\ to 3200 \AA. We provide
access to the UV light curves and other basic information about the observations
in tabular and graphical form via the Internet at
http://www.chara.gsu.edu/PEGA/IUE/.

\end{abstract}

\keywords{galaxies: Seyfert -- ultraviolet: galaxies}

\section{Introduction}

Seyfert galaxies harbor Active Galactic Nuclei (AGN) with typical redshifts 
$z \lesssim$ 0.1 and moderate luminosities that vary over a range of 
time scales from days to years and over a factor of $\sim$10 in amplitude. 
AGN have supermassive black holes consuming nearby gas and/or stars; 
the accretion disk surrounding the black hole is presumably the source for 
their high apparent luminosity and continuum variability. Seyfert galaxies 
are classified as types 1 and 2. Seyfert 1 galaxies have narrow ($\sim$500 
km s$^{-1}$ full width half maximum [FWHM]) forbidden and permitted 
emission lines and broad ($>$ 1000 km FWHM) permitted lines, while Seyfert 2 
have only narrow permitted and forbidden emission lines. (Khachikian \& 
Weedman 1974). Ultraviolet spectra of Seyfert galaxies are notable for their
strong continuum variability, which is not as pronounced in the visual regime
(e.g., NGC 4151, Kaspi et al. 1996).

The {\it International Ultraviolet Explorer (IUE)} began monitoring AGN in 1978 
and collected a large number of UV spectra through 1995, while the Faint
Object Spectrograph (FOS) on board the {\it HST} collected spectra from 1990 
through 1997.  Although the origin of the UV continuum is still elusive, it is 
currently believed that it is radiated from the accretion disk. Comparisons of 
variability in the UV with other regions of the spectrum places important 
constraints on the physics involved (e.g.,Nandra \& Papadakis 2001). Thus, it 
is important to have a comprehensive and uniform database of UV light curves 
for comparison with continuum observations at other wavelengths. 

From the massive collection of {\it IUE} and FOS spectra, we have compiled
continuum light curves and created a database that is internet accessible. Our 
list of targets includes any object that has been designated a Seyfert Galaxy by the 
observing astronomer for any observation with either {\it IUE} or {\it HST}. Also, we 
have limited the selection list via redshift, including only objects with a redshift 
z $<$ 0.2. Our database is the first effort to make AGN UV light curves readily 
available via the World Wide Web. The only previous effort was an atlas of {\it IUE} 
spectra of Seyfert Galaxies observed prior to 1991 January 1 (Courvoisier \& 
Paltani 1992). The database we have created is inclusive of all {\it IUE} and FOS 
spectra through their final years, and has the capability to receive further 
observations provided by sources such as the Goddard High Resolution Spectrograph 
(GHRS), Space Telescope Imaging Spectrograph (STIS), and the {\it Hopkins Ultraviolet 
Telescope (HUT)}. This information should be helpful for a number of studies, 
including those that require 1) the history of the UV continuum variations for an 
individual Seyfert galaxy, 2) a comparison of UV fluxes with measurements in other 
regimes of the electromagnetic spectrum for any given epoch, and 3) detailed 
statistical analyses (e.g., cross-correlation,structure functions) of the UV 
continuum properties of AGN.

\section{Observations}

\subsection{IUE}

{\it IUE} was launched on 1978 January 26 and operated successfully until 30 
September 1996, when it was decommissioned. {\it IUE} had the capability to 
perform spectroscopy at two different resolutions. The high resolution mode 
operated at a spectral resolution of 0.1 \AA\ to 0.3 \AA\ (FWHM), while the low 
resolution mode performed at 6 \AA\ to 7 \AA\ (FWHM).  {\it IUE} had two
apertures available for spectroscopy: large (10 x 20
arcsec) and small (3 arcsec in diameter). Due to the faintness of
Seyfert galaxies, the high-resolution mode is unsuitable for continuum studies.
Observations through the small aperture are unsuitable for absolute photometry
due to the large ($\sim$ 50\%) and variable light loss. Thus we used only
large-aperture, low-dispersion spectra for the light curves. {\it IUE} utilized
a total of four cameras: 
the Long Wavelength Prime and Redundant Cameras (LWP and LWR) and the Short 
Wavelength Prime and Redundant Cameras (SWP and SWR), which operated at 1850 \AA\ 
to 3200 \AA\ and 1150 \AA\ to 2000 \AA\ respectively.  The SWR Camera was only 
used for a handful of observations early on, and no useful data on Seyfert galaxies 
were obtained. Both the LWP and LWR cameras were used to make observations over 
the lifetime of {\it IUE}.  During {\it IUE}'s early life the LWR camera
dominated the observing; during the later years the LWP camera become the
dominant camera due to a flare that developed in the LWR
\footnote {See http://archive.stsci.edu/iue/ for detailed information on the
{\it IUE} telescope and instruments}.

\subsection{FOS}

The FOS was one of the initial spectrographs installed 
on the {\it HST}, which was launched in 1990. It remained on board Hubble until 
1997. There are two versions of the FOS data, pre-COSTAR and post-COSTAR,
corresponding to data obtained before and after the optics were repaired on
{\it HST} in 1994 January. The optics correction caused a slight change in the
aperture sizes, but there were no major effects on the quality of the data. 
We chose to use data from all apertures, since aperture corrections for the FOS
are highly reliable. The FOS had two spectral resolutions available, high
($\lambda$/$\Delta$$\lambda$
$\sim$1300) and low ($\lambda$/$\Delta$$\lambda$ $\sim$ 250). For our sample we
chose to use the high-resolution grating due to the short exposure times and
resulting poor signal-to-noise in the low resolution spectra. Although 
the FOS gratings covered both the UV and visible regimes, we chose to use only 
the UV data, from gratings G130H, G190H and G270H with wavelength ranges 
spanning 1000 -- 1700 \AA\, 1700 -- 2200 \AA\ and 2200 -- 3000 \AA\ respectively
\footnote {See http://www.stecf.org/poa/FOS/ for detailed information on the
FOS}.

\section{Data Reduction and Analysis}

For our purposes, every available spectrum was obtained using the MAST
interface. The first objective was to create a single averaged spectrum per AGN
from a short list of viable candidates, to identify suitable continuum regions. 
We chose NGC 5548, NGC 4151, Fairall 9 and Mrk 509,  primarily due to their
numerous {\it IUE} observations (367, 984, 233, and 97 observations,
respectively). To view the common emission lines, the average spectra were
de-redshifted and plotted together, as shown in Figures 1 and 2.  We determined
from the figures where continuum flux measurements were possible with little
chance of an emission or absorption line altering the flux measurements. The
best locations for the SWP camera measurements fell at 1355, 1720, and 1810 \AA\
with bin sizes of 30, 30, and 50 \AA, whereas the LWP/LWR camera bins fell at
2200, 2400 and 2740 \AA\ with bin sizes of 50, 60 and 30 \AA, all in the rest 
frame of the galaxy. Note that fluxes in the 2200 \AA\ bin are rather noisy 
compared to the other bins, due to the low sensitivities of the cameras in
this region.

Our next step was to remove flawed spectra from the sample.  Several reasons arose 
to eliminate an {\it IUE} spectrum from the light curve list. One such reason
was overexposure. The {\it IUE} cameras had a limited dynamic range, such that
the raw counts (in data numbers [DN]) could not exceed the value of 255.
Although in most cases only the emission lines were overexposed, we decided to
be cautious and not use overexposed images for our continuum measurements.
Another problem that appeared was heavy background noise during periods of {\it
IUE}'s orbit.  As {\it IUE} orbited the Earth it would daily fall into the Van
Allen belts. These events would send background counts up to high levels,
thereby making the continuum fluxes very noisy. Low to moderate 
background values turned out to be between 50 and 100 DN, above 100 DN was
considered a high value, and above 200 DN as unusable. To set a standard limit,
we have omitted any spectra with background $>$ 100 DN.  During
{\it IUE}'s later years, sunlight leaked into the telescope and added an
extended source of light at wavelengths longer than 2500 \AA. All spectra
observed after 1990 that exhibited an unusually large increase in continuum
slope in the LWR and LWP images due to this effect were excluded from our light
curves. Other less prominent events affected the {\it IUE} spectra, but were
either rare or did not affect the continuum flux.  We provide in our
web site a list of spectra for each object that were removed from our
measurements along with a key of explanations as to what events occurred to
exclude the spectra.

Once we established a quality standard, we shifted our predetermined wavelength bins by 1+z 
to measure the fluxes in the object's frame. For example, in NGC 4151 (z $=$ 0.0033) the SWP
bin positions shifted to 1359, 1725, and 1816 \AA.  When redshifting a bin near the edge 
of a spectrum, the 1810 bin in the SWP and the 2740 bin in the LWP camera would occasionally
fall off the spectrum and yield a null result. This occurance is the reason for a redshift 
limit of z $<$ 0.2 in our selection list.

As discussed previously, the FOS had three fixed grating ranges available. This 
meant that unlike the {\it IUE} dataset, there was typically one bin in the
G130H (the 1355 \AA\ bin), three bins in the G190H (the 1720, 1810 and 2200 \AA\ bins) 
and two bins in the G270H (the 2400 and  2740 \AA\ bins) spectra.  When examining 
light curves comprised of data from each source, it should be noted that the quality
of spectra for measurements in the 1355 and the 1720 bins on the FOS could be
vastly different, yet with {\it IUE} observations these two bins were obtained
at the same time and with the same camera and thus less likely to contain
deviations from one another. 

Once the measurements were made, we plotted the average flux in each bin versus the 
Julian Date of the midpoint of each exposure, calculated from the start
time of the observation and the exposure length, which were taken from the file
header provided by MAST. In order to extend the time span in which objects
were monitored, we combined the light curves from FOS and {\it IUE}. An example
of these results is displayed for NGC 4151 in Figure 3, along with a "zoomed in"
section in Figure 4. In these light curves, it is obvious that over long time
frames (years) the flux can change by a factor of 10, as previously mentioned. 
In Figure 4 it can be seen that small flux changes occurred over the course of
three days time in the intense monitoring campaign during this time period. This
is one of the few occasions in which the observations did not undersample
the continuum variations (Crenshaw et al. 1996).

Continuum fluxes at the level of $\sim$2 $\times$ 10$^{-15}$ ergs s$^{-1}$
cm$^{-2}$ \AA$^{-1}$ or smaller are unreliable, due to camera artifacts and
uncertain background levels in the {\it IUE} cameras (Crenshaw et al.
1990). Note that extended continuum sources, such as starbursts or scattering
regions often found in Seyfert 2 galaxies, could show apparent flux variability 
due to the different aperture sizes used by {\it IUE} and the FOS.

\section{Error Analysis}

In order to determine the uncertainties in the {\it IUE} fluxes, we began by
establishing the average of the continuum in each bin, illustrated by the
horizontal line in Figure 5, and then measured the standard deviation ($\sigma$)
of the points within the designated bins. This method for {\it IUE} error 
determination is known to overestimate the uncertainties (Clavel et al. 1991).
Because the {\it IUE} cameras were not photon counters and the spectra were
highly oversampled, there are approximately 4 spectral points per wavelength
resolution element.

For our measurements we concluded that a method of adjusting the error bars similar 
to the method of Clavel et al. (1991) is optimal. On time scales less than
1 -- 2 days, previous studies have shown that the UV continuum variations
in Seyfert galaxies are small (see Figure 4). Thus any apparent
continuum variablity exhibited on very short time scales in the {\it
IUE} data is likely dominated by photon and/or instrument noise.
To test this notion, we did combinatorics of all points in
each light curve, for several objects that contained at least tens of
observations, to find the fractional change in flux ($\Delta$f) and time
($\Delta$t) between any two points (it should be noted that this is similar to a
structure function method). Figure 6 shows a plot of $\Delta$f against
($\Delta$t) for each pair of points in Figure 3. As expected, as
the time interval grows, the maximum change in flux grows. The
leveling off of points on time intervals less than $\sim$1 day
confirms that there is no significant excess variability above the noise level
(with the exception of a few points) on short time scales.
 
We took measurements of the light curve points contained within a rectangle between 
log($\Delta$t/days) of -1.5 and -0.5 and between log($\Delta$f) of 0.0 and 0.15 in
Figure 6 to characterize the noise. We define the reproducibility R as the
average of the fractional variations in the time bin. R represents an upper
limit to the flux uncertainty (on average), since we cannot rule out the 
possibility of small variations on these short time scales. The mean measured
error (EMean) is given by the average of all of the uncertainties ($\sigma$s)
in this time interval. The overestimation of the error is given by the ratio of
the mean measured error EMean and the reproducibility (R). Thus, if we
divide our individual $\sigma$'s by this ratio, we obtain an estimate of the
noise in each measurement, but scaled appropriately. 

We repeated the above procedure for a number of well-observed Seyfert galaxies.
In Tables 1, 2, and 3 we present the necessary data to determine the
overestimation of the error for each camera. MaxJD and MaxFl are the
logarithmic limits we imposed on the region of the plot of $\Delta$f vs.
$\Delta$jd to determine the reproducibility. Ratio is the average error
(EMean) divided by the reproducibility (R), or, in other words, the factor by
which the original errors have been overestimated. The reciprocal of the ratio 
averaged over all objects gives the final scaling factor for theoriginal errors. 
The average ratios are 1.69 $\pm$ 0.34, 2.40 $\pm$ 0.77, and 1.75 $\pm$ 0.31 for 
the SWP, LWR, and LWP cameras respectively. For the FOS data, in the few 
instances where we have multiple observations of an object over short time scales, 
we see no evidence that the error bars require additional scaling.

\section{Database and Website}

We have created a web site to make our UV light curves of Seyfert
galaxies available to the community. Users can access our web site at
www.chara.gsu.edu/PEGA/IUE. We chose to create a MySQL database to store the
data and build our web site with Perl CGI scripts.  The opening page is a
tabular list of the objects for which we have data, sorted by Right Ascension
(see Figure 7). Users can use their web browser's "find" function to locate
an object by name. The main page presents not only the list of objects and
positions, but their respective redshifts and a link to the light curve for
each bin that was measured (the central position of the bin is given in the
observed frame). For each link, we provide an on-the-fly light curve generator
(Figure 8). The user is able to adjust the time frame or the flux range
within the window to allow for a zoom.  For each light curve, users can
left click on any point to access the observing information for that point.
We also provide a link to the information for all observations in the band
pass, as well as the light curve in tabular form (Julian Date, fluxes, and
scaled uncertainties) for downloading, as shown in Figure 9. This list
also gives the observations that were removed from the light curve, if any. From
this page, users are also able to access previews of the original spectra on the
MAST web site. In the future, we hope to add points to the light curves from
other UV spectrographs, such as GHRS, STIS, and {\it HUT}.

\acknowledgments

This research has made use of the NASA/IPAC Extragalactic Database (NED) which 
is operated by the Jet Propulsion Laboratory, California Institute of 
Technology, under contract with the National Aeronautics and Space 
Administration. All of the data presented in this paper were obtained 
from the Multimission Archive at the Space Telescope Science Institute (MAST). 
Support for MAST for non-HST data is provided by the NASA Office of Space 
Science via grant NAG5-7584 and by other grants and contracts.

\clearpage

\figcaption[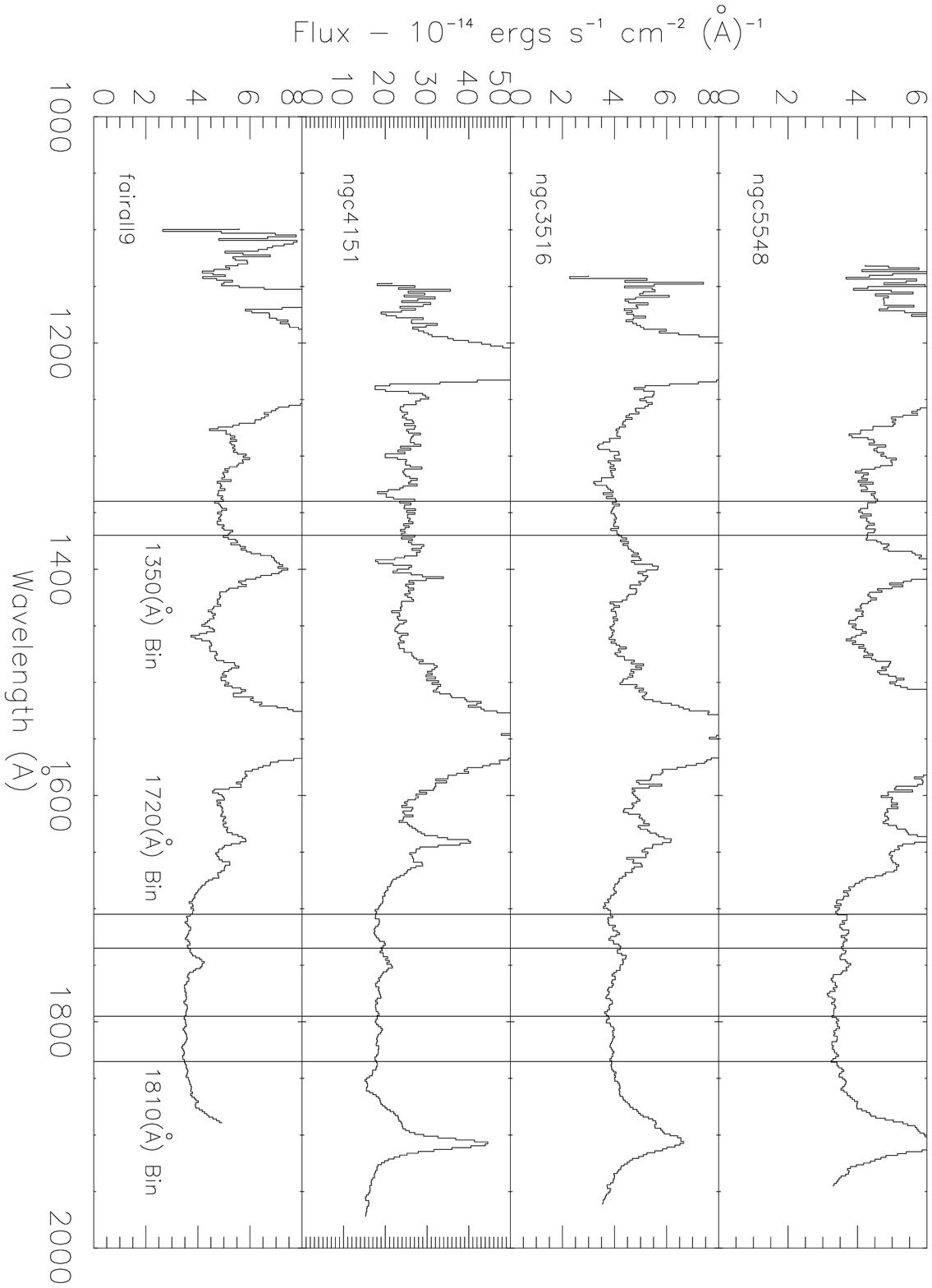]{Plots of {\it IUE} SWP spectra for four frequently
observed Seyfert galaxies in each object's rest frame to determine the best
locations and sizes for bin measurements. These are average spectra from all of
the good-quality spectra provided by MAST. The bin locations and widths used for
determining the light curves are shown.}

\figcaption[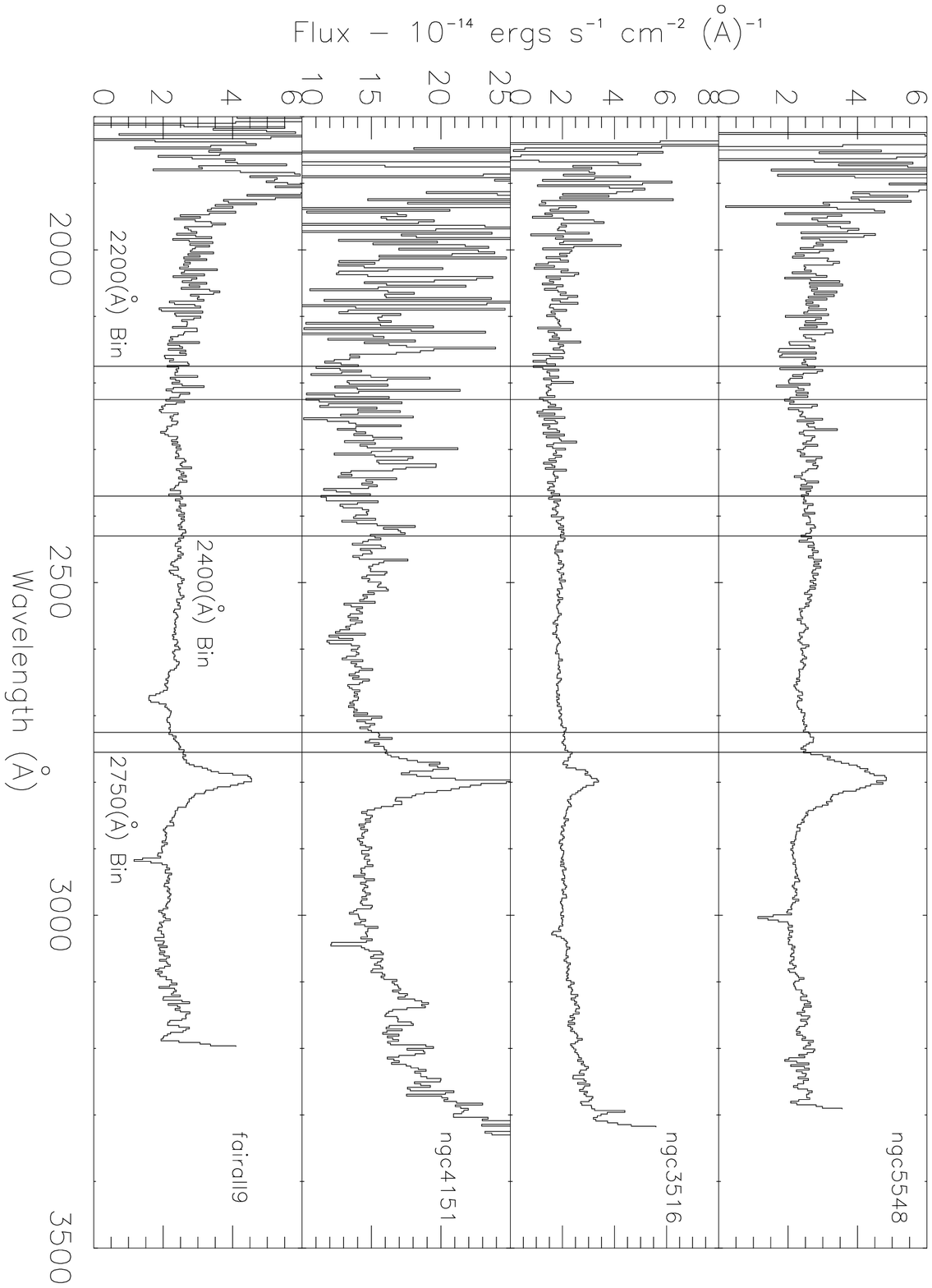]{Plot of average LWP spectra for the same four Seyfert galaxies
as in Figure 1.  Bin locations and widths are indicated.}

\figcaption[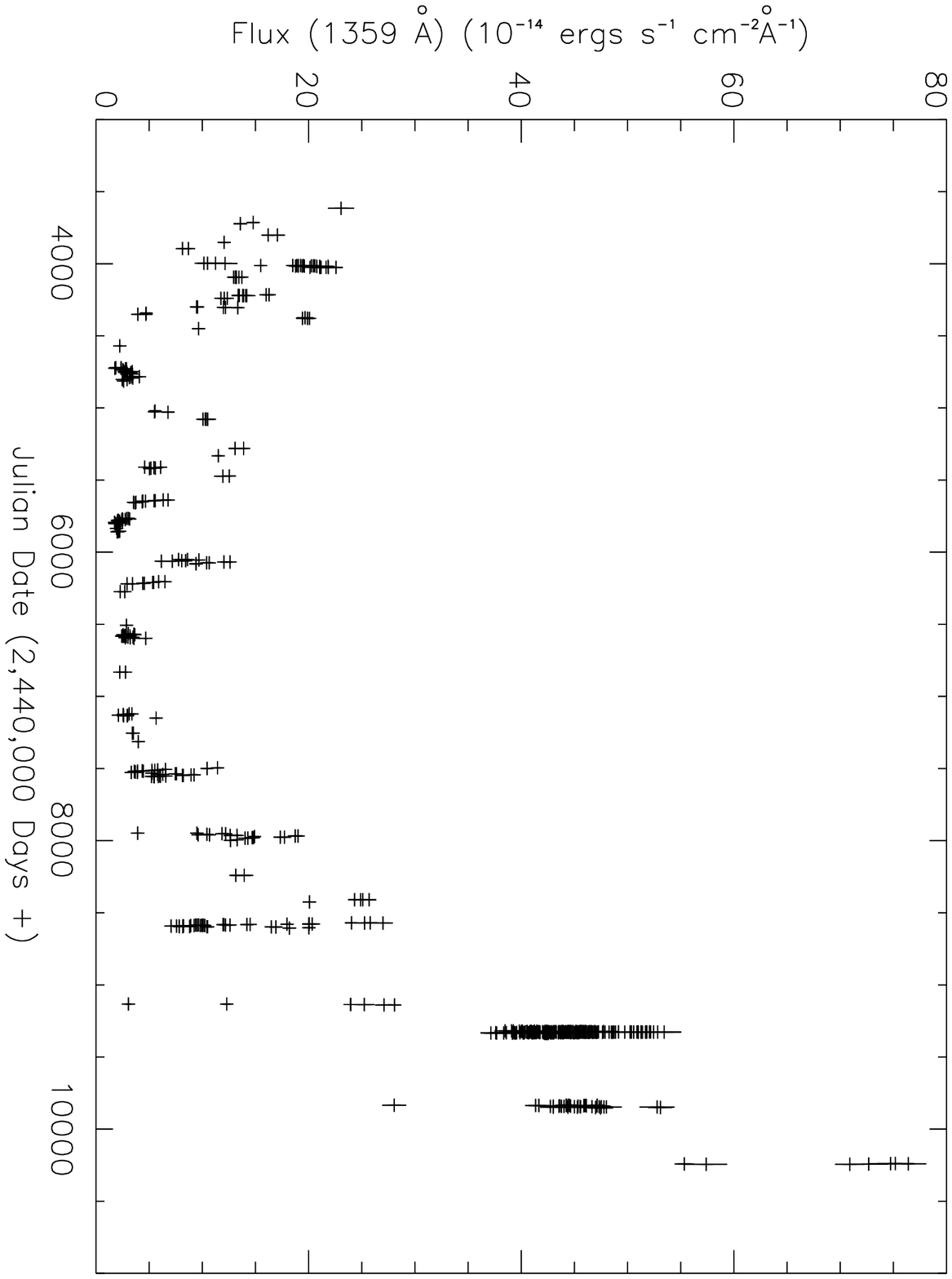]{Example light curve plot of the 1355 bin for NGC 4151.
This particular light curve shows an excellent example of large scale changes in
flux over significant periods of time.}

\figcaption[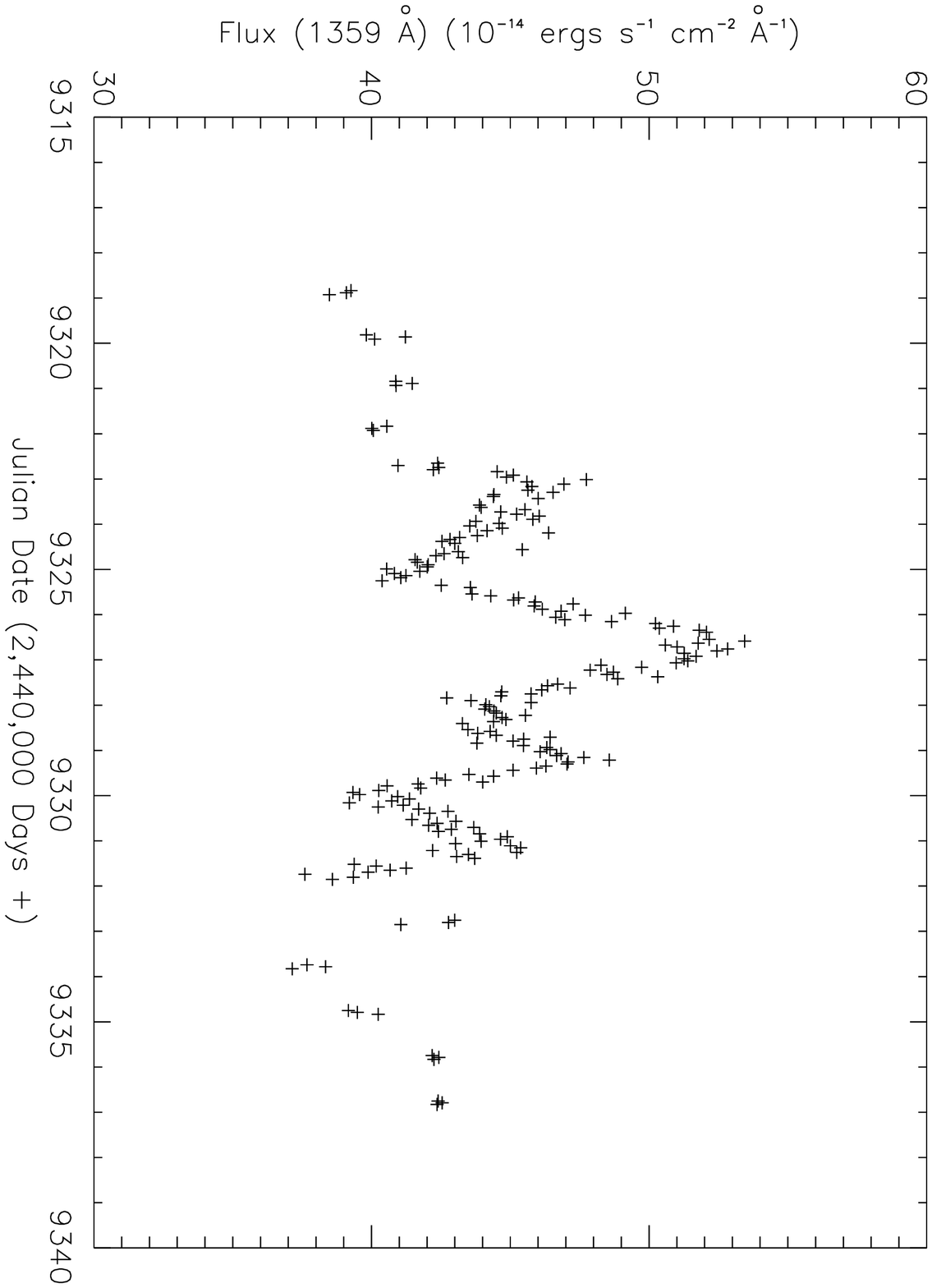]{Zoomed version of Figure 3. Error bars are not shown for
clarity.}

\figcaption[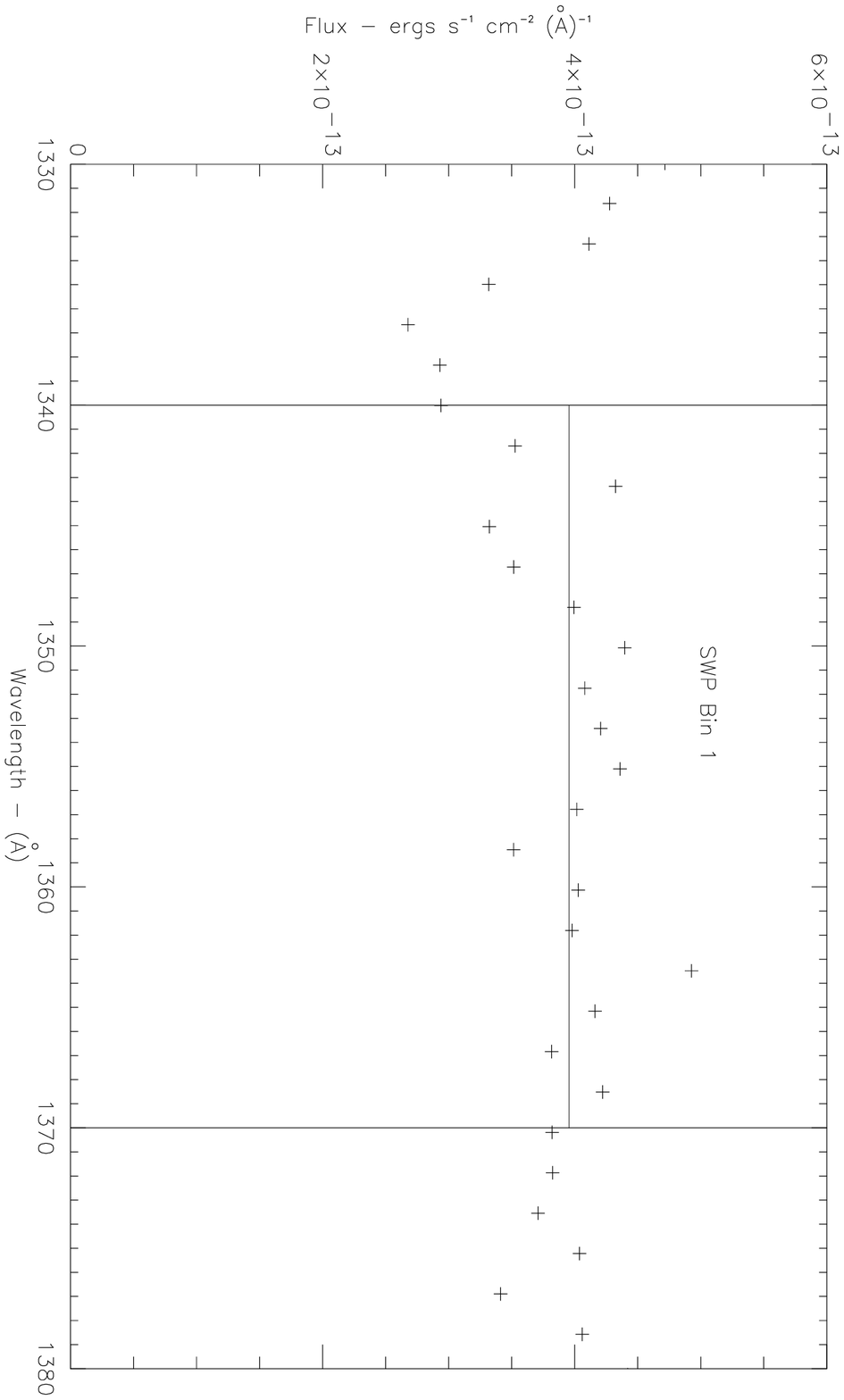]{Example of the 1355 bin with the SWP camera for NGC 4151, 
zoomed to show the average flux within the bin. The error calculated is 
the standard deviation of the observed points from that average flux.}

\figcaption[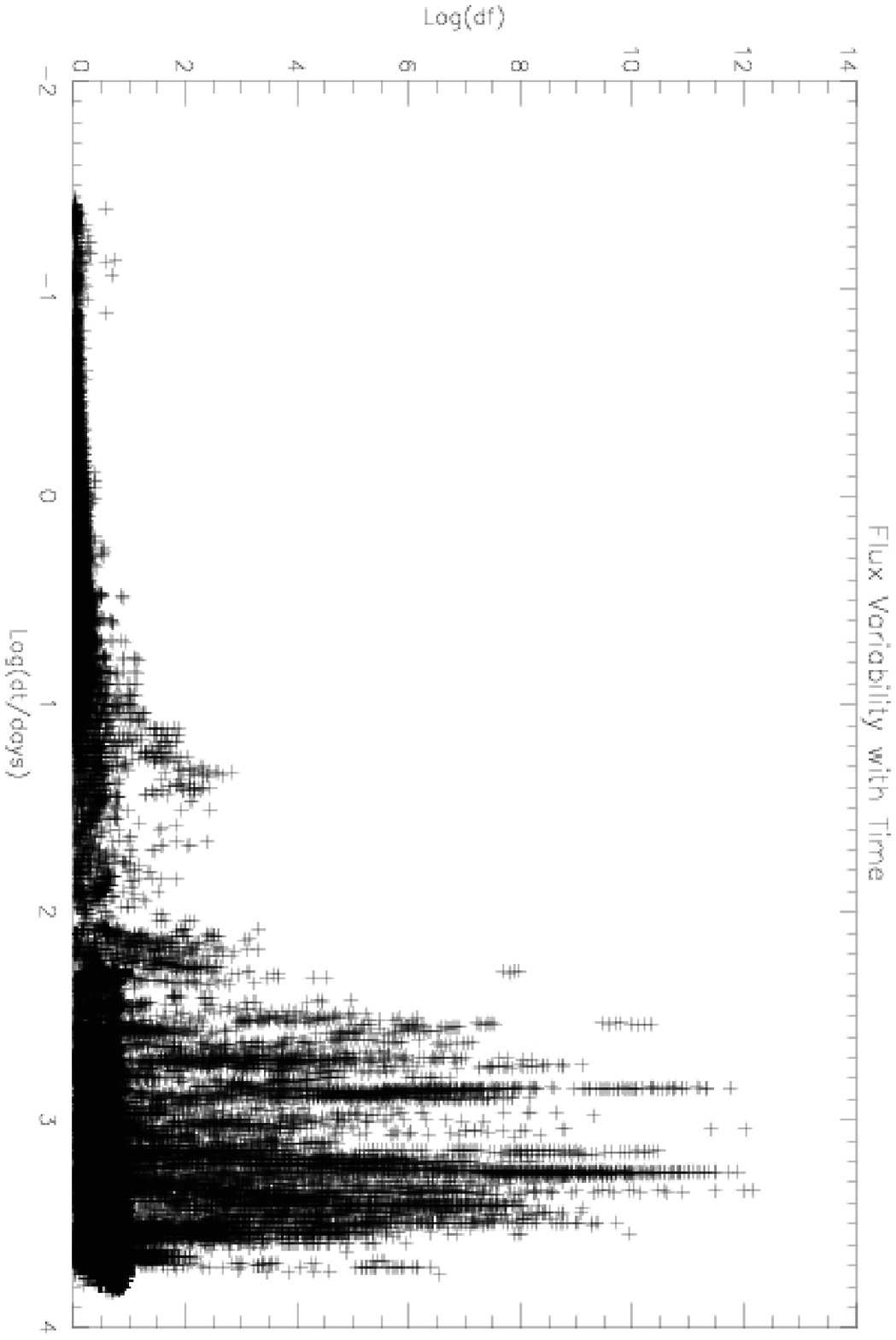]{Plot of the change in flux versus the change in time for NGC 4151.}

\figcaption[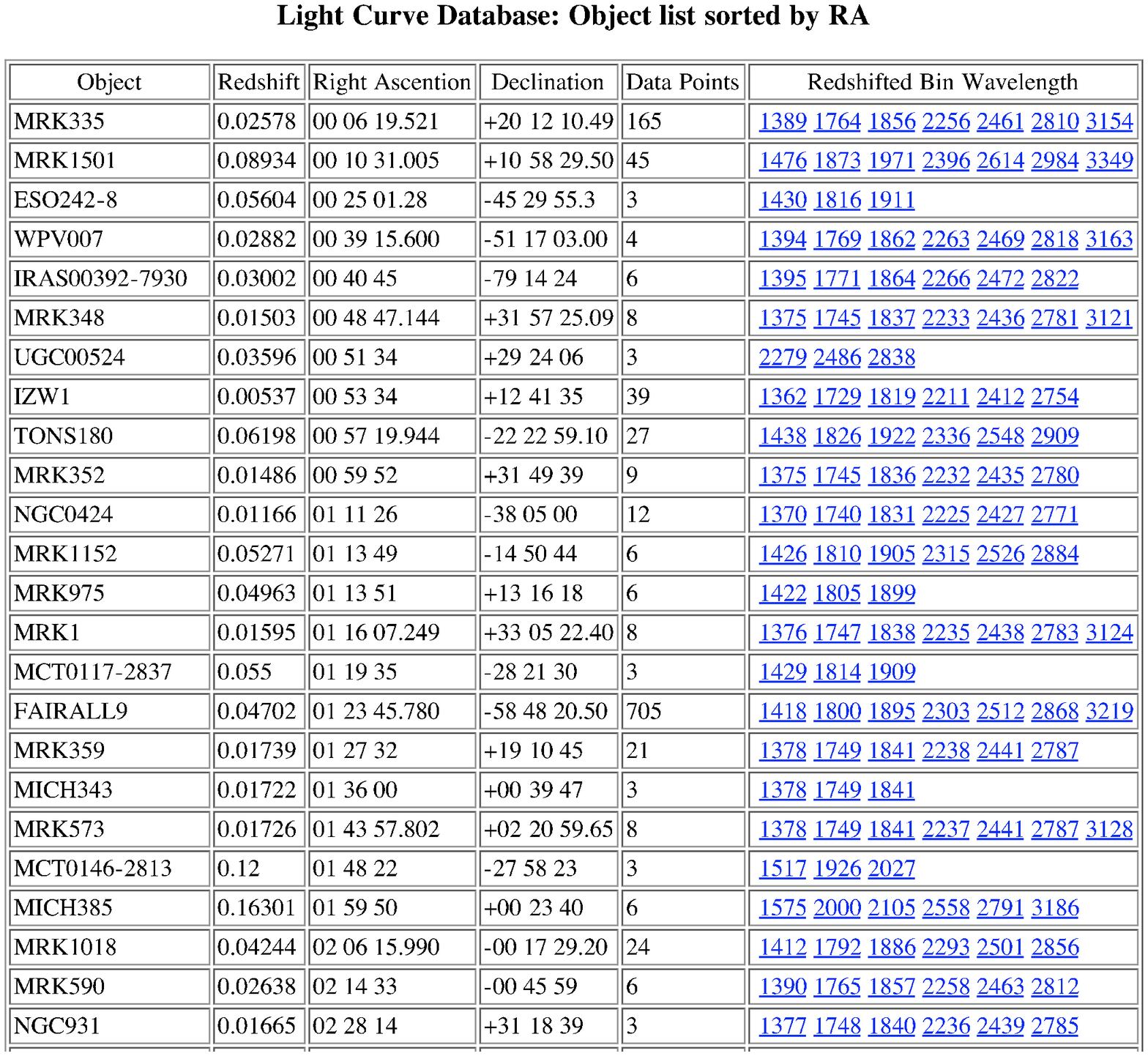]{Sample image of web site home page.}

\figcaption[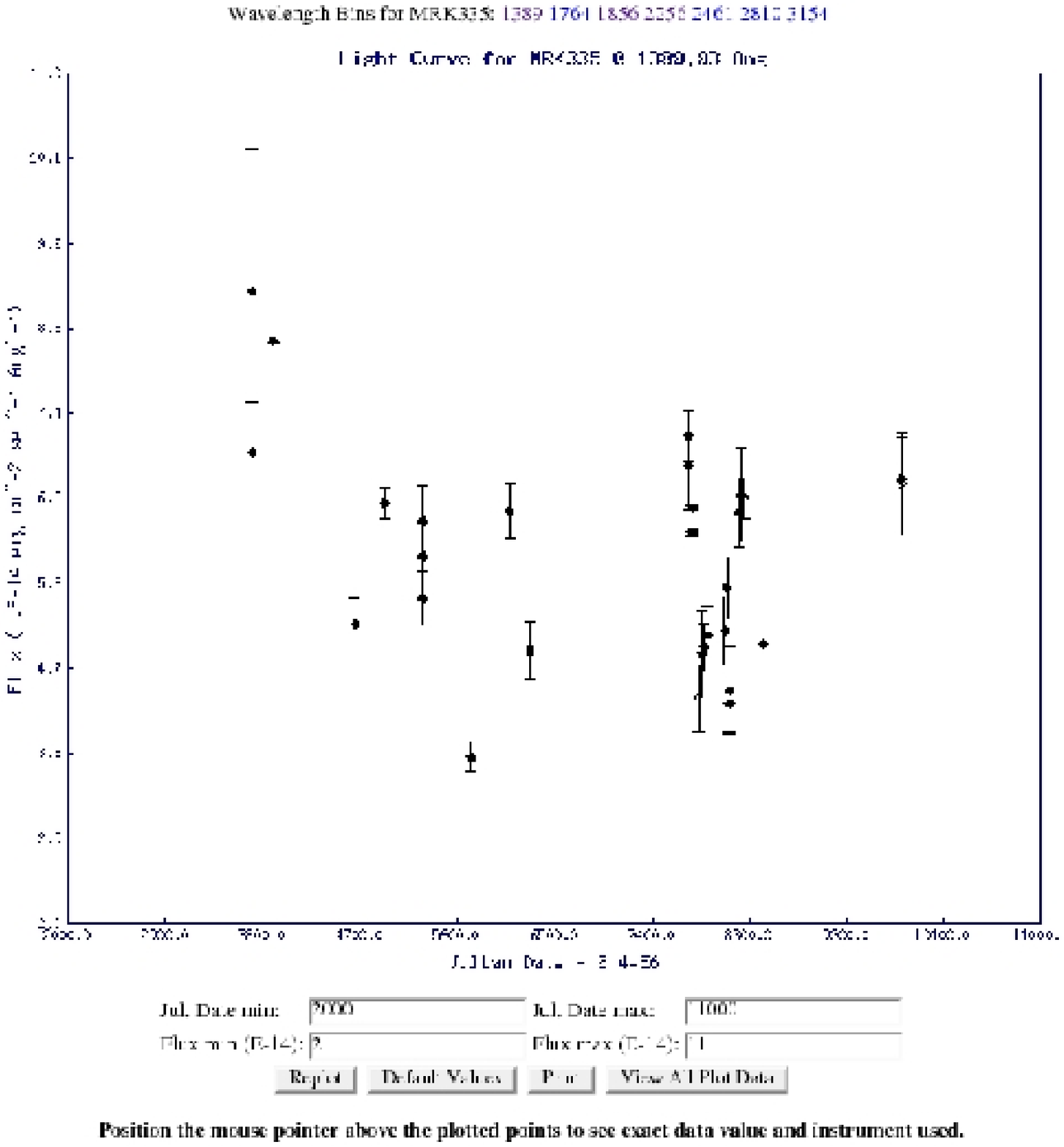]{Sample image of the light curve on-the-fly generator.}

\figcaption[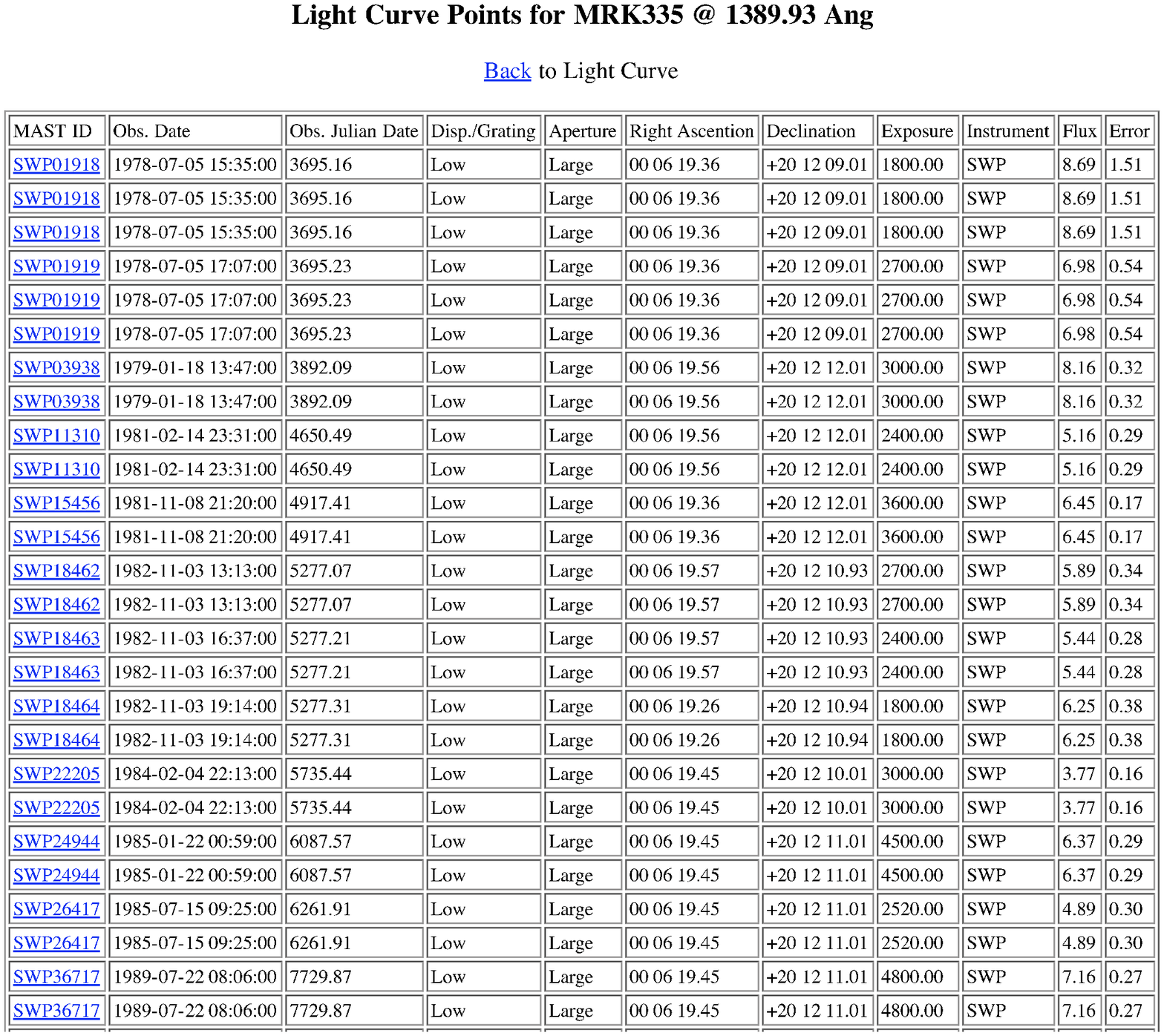]{Sample of the light curve information page.}

\clearpage
\begin{figure}
\plotone{fg1.eps}
\\Fig.~1.
\end{figure}

\clearpage
\begin{figure}
\plotone{fg2.eps}
\\Fig.~2.
\end{figure}

\clearpage
\begin{figure}
\plotone{fg3.eps}
\\Fig.~3.
\end{figure}

\clearpage
\begin{figure}
\plotone{fg4.eps}
\\Fig.~4.
\end{figure}

\clearpage
\begin{figure}
\plotone{fg5.eps}
\\Fig.~5.
\end{figure}

\clearpage
\begin{figure}
\plotone{fg6.eps}
\\Fig.~6.
\end{figure}

\clearpage
\begin{figure}
\plotone{fg7.eps}
\\Fig.~7.
\end{figure}

\clearpage
\begin{figure}
\plotone{fg8.eps}
\\Fig.~8.
\end{figure}

\clearpage
\begin{figure}
\plotone{fg9.eps}
\\Fig.~9.
\end{figure}

\newpage
\begin{deluxetable}{ccccccc}
\tablecolumns{7}
\footnotesize
\tablecaption{Error Analysis of {\it IUE} data for Seyfert Galaxies (SWP)}
\tablehead{
\colhead{Object} &
\colhead{$\sigma$} &
\colhead{Log Max JD} &
\colhead{Log Max Fl} &
\colhead{R} &
\colhead{E Mean} &
\colhead{Ratio}
}
\startdata
NGC4151     & 0.026 & -0.50 & 0.15 & 0.032 & 0.067 & 2.06 \\
NGC7469     & 0.037 & -0.50 & 0.20 & 0.045 & 0.076 & 1.67 \\
NGC3783     & 0.052 & 0.00 & 0.30 & 0.054 & 0.073 & 1.36 \\
NGC3516     & 0.039 & 0.20 & 0.15 & 0.046 & 0.081 & 1.76 \\
NGC5548     & 0.041 & 0.25 & 0.15 & 0.048 & 0.085 & 1.78 \\
Fairall9    & 0.040 & 0.25 & 0.15 & 0.057 & 0.099 & 1.74 \\
3c390.3     & 0.129 & 1.50 & 0.50 & 0.222 & 0.272 & 1.22 \\
Mrk509      & 0.031 & 0.75 & 0.20 & 0.040 & 0.080 & 1.99 \\
Mrk335      & 0.076 & 1.00 & 0.30 & 0.091 & 0.099 & 1.09 \\
NGC4593     & 0.107 & 1.00 & 0.50 & 0.129 & 0.115 & 0.89 \\
Rxj 2304.7-0841 & 0.096 & 1.50 & 0.50 & 0.114 & 0.150 & 1.32 \\
\enddata
\normalsize
\end{deluxetable}

\begin{deluxetable}{ccccccc}
\tablecolumns{7}
\footnotesize
\tablecaption{Error Analysis of {\it IUE} data for Seyfert Galaxies (LWR)}
\tablehead{
\colhead{Object} &
\colhead{$\sigma$} &
\colhead{Log Max JD} &
\colhead{Log Max Fl} &
\colhead{R} &
\colhead{E Mean} &
\colhead{Ratio}
}
\startdata
NGC4151     & 0.022 & -0.50 & 0.15 & 0.030 & 0.103 & 3.43 \\
NGC7469     & 0.007 & 1.50 & 0.20 & 0.054 & 0.082 & 1.52 \\
NGC3783     & 0.474 & 0.50 & 0.15 & 0.072 & 0.109 & 1.51 \\
NGC3516     & 0.555 & 2.30 & 0.20 & 0.078 & 0.102 & 1.31 \\
NGC5548     & 0.031 & 1.50 & 0.15 & 0.064 & 0.164 & 2.56 \\
Fairall9    & 0.022 & 1.00 & 0.10 & 0.034 & 0.099 & 2.91 \\
3c390.3     & 0.057 & 2.50 & 0.20 & 0.073 & 0.214 & 2.93 \\
Rxj 2304.7-0841 & 0.032 & 2.00 & 0.20 & 0.037 & 0.111 & 3.00 \\
\enddata
\normalsize
\end{deluxetable}

\newpage
\begin{deluxetable}{ccccccc}
\tablecolumns{7}
\footnotesize
\tablecaption{Error Analysis of {\it IUE} data for Seyfert Galaxies (LWP)}
\tablehead{
\colhead{Object} &
\colhead{$\sigma$} &
\colhead{Log Max JD} &
\colhead{Log Max Fl} &
\colhead{R} &
\colhead{E Mean} &
\colhead{Ratio}
}
\startdata
NGC4151     & 0.022 & -1.00 & 0.15 & 0.028 & 0.050 & 1.78 \\
NGC3783     & 0.039 & 0.00 & 0.15 & 0.032 & 0.043 & 1.34 \\
NGC5548     & 0.062 & 0.50 & 0.25 & 0.064 & 0.088 & 1.38 \\
Fairall9    & 0.027 & 0.00 & 1.00 & 0.028 & 0.058 & 2.07 \\
3c390.3     & 0.027 & 1.70 & 1.20 & 0.056 & 0.120 & 2.14 \\
Mrk335      & 0.036 & 1.00 & 0.10 & 0.049 & 0.087 & 1.78 \\
\enddata
\normalsize
\end{deluxetable}

\end{document}